\documentclass[11pt, a4]{article}

\usepackage[toc,page]{appendix} 
\usepackage[breaklinks]{hyperref}
\usepackage[T1]{fontenc} 
\usepackage[latin1]{inputenc}
\usepackage[top=1in, bottom=1in, left=1in, right=1in]{geometry}
\usepackage{aeguill}
\usepackage{tikz}
\usepackage{amsfonts}
\usepackage{amsmath}
\usepackage{amssymb}
\usepackage{amssymb,stmaryrd}
\usepackage{gnuplot-lua-tikz}
\usepackage{tikz-uml}
\usepackage{subfigure}
\usepackage{multirow}
\usepackage{garamond}
\usepackage[sort&compress]{natbib}
\usepackage[]{url,hyperref}
\usetikzlibrary{arrows,automata,shapes}

\usepackage{listings,color}

\definecolor{gray}{rgb}{0.4,0.4,0.4}
\definecolor{darkblue}{rgb}{0.0,0.0,0.6}
\definecolor{cyan}{rgb}{0.0,0.6,0.6}
 
\lstdefinelanguage{XML}
{
  basicstyle=\ttfamily\tiny\color{darkblue}\bfseries,
  breaklines = true,
  morestring=[b]",
  morestring=[s]{>}{<},
  morecomment=[s]{<?}{?>},
  stringstyle=\color{black},
  identifierstyle=\color{darkblue},
  keywordstyle=\color{cyan},
  morekeywords={xmlns,version,type}% list your attributes here
}

\begin{document}

\garamond

\title{Dynamic Hybrid Traffic Flow Modeling\\~\\{\small CISIT Phase 5 --- ISART Project}\\{\small Scientific report 2013}~\\~\\}

\author{Hassane Abouaïssa, Yoann Kubera, Gildas Morvan\\~\\
{\small \url{http://www.lgi2a.univ-artois.fr/~morvan/}}\\{\small\href{mailto:gildas.morvan@univ-artois.fr}{gildas.morvan@univ-artois.fr}}\\~\\
Univ Lille Nord de France, F-59000 Lille, France\\UArtois, LGI2A, F-62400, Béthune, France\\~\\~\\~\\
\includegraphics[width=3cm]{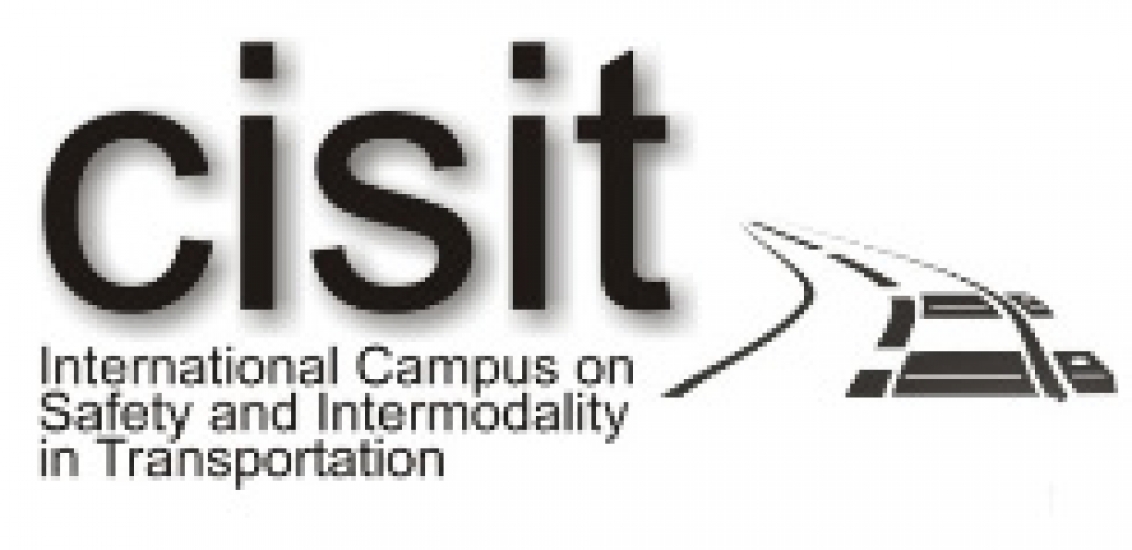}~~~\includegraphics[width=3cm]{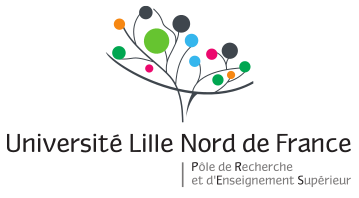}~~~\includegraphics[width=3cm]{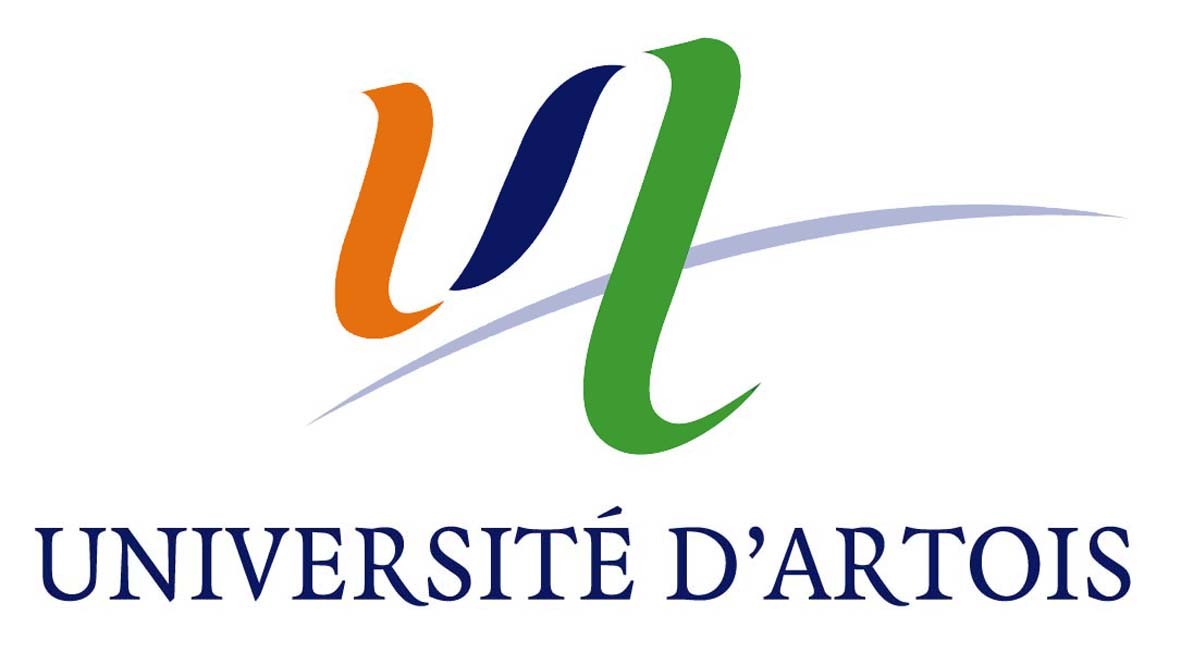}~~~\includegraphics[width=2.5cm]{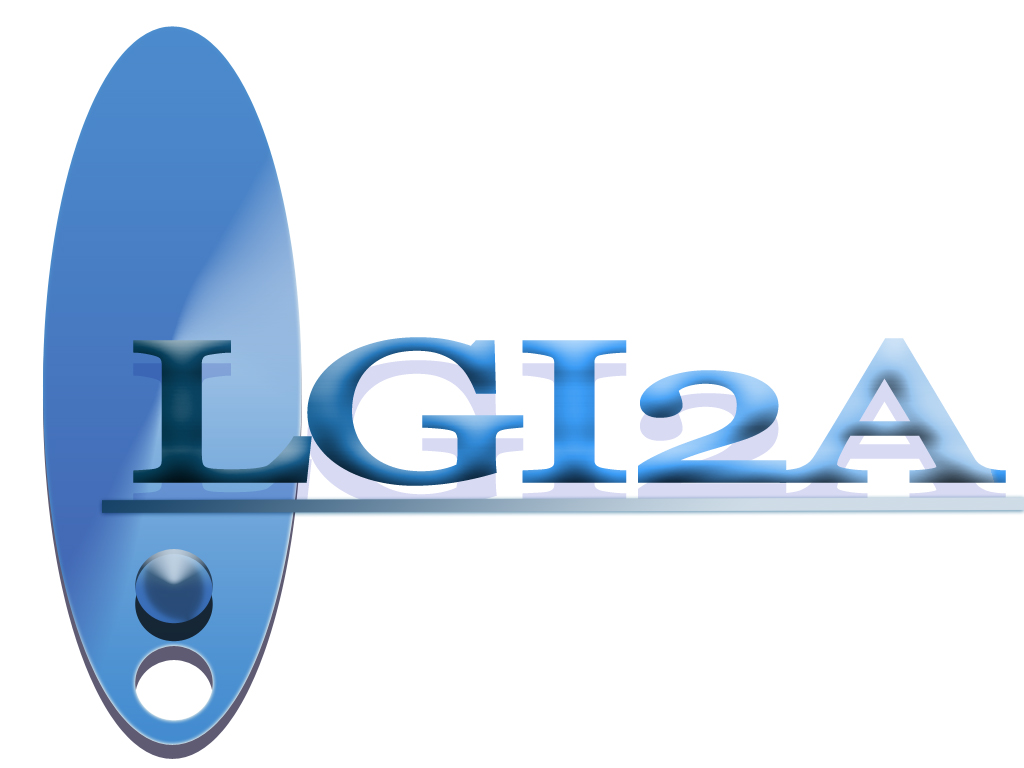}\\~\\~\\~\\~\\~\\~\\~\\~\\
}

\date{}

\clearpage
\maketitle
\thispagestyle{empty}

\setcounter{footnote}{0}

\paragraph{Acknowledgments}{\small This work has been supported by International Campus on 
Safety and Intermodality in Transportation (CISIT), the Nord-Pas-de-Calais 
Region, the European Community, the Regional Delegation for Research and 
Technology, the Ministry of Higher Education and Research, and the National 
Center for Scientific Research (CNRS).}

\pagebreak

%\begin{scriptsize}
 \tableofcontents
%\end{scriptsize}

\section*{Résumé en français}

{\footnotesize La simulation du trafic  routier sur des réseaux de grande échelle est un problème compliqué car il suppose d'intégrer dans un même modèle différentes approches. Ainsi, les sections autoroutières sont généralement représentées à l'aide de modèles macroscopiques alors que pour les sections urbaines, des modèles microscopiques sont utilisés. De manière générale, les modèles microscopiques sont intéressants lorsque les interactions entre véhicules, ainsi que la topologie du réseau deviennent complexes.

Des modèles intégrant ces différents niveaux de représentation sont généralement qualifiés d'hybrides. Par ailleurs, ils sont généralement "statiques"~: à chaque portion du réseau est associée une représentation unique qui ne changera pas au cours de la simulation. Afin de palier cette limitation, Nous avons  débuté en 2013 dans le cadre du projet ISART le développement d'un simulateur multi-agent multi-niveaux de flux de trafic routier nommé JAM-FREE permettant :
\begin{itemize}
\item de simuler des réseaux routier de grande taille efficacement en utilisant la technique du niveau de détail dynamique,
\item de tester de nouveaux algorithmes de régulation, observation et routage.
\end{itemize}

Ce simulateur repose sur un framework de modélisation et de simulation multi-agents multi-niveaux nommé SIMILAR (\textbf{SI}mulations with \textbf{M}ult\textbf{I}-\textbf{L}evel \textbf{A}gents and \textbf{R}eactions), implémenté en Java~\citep{Morvan:2014,Morvan:2014a}, distribué prochainement sous licence libre.

Dans ce rapport, nous présentons ces résultats scientifiques ainsi que les publications associées.}

\section{Introduction}
\label{Introduction}

This paper reports the works conducted by Hassane Abouaïssa\footnote{\url{http://www.lgi2a.univ-artois.fr/spip/spip.php?article5&idpersonne=5}}, Yoann Kubera\footnote{\url{http://www.yoannkubera.net/}} and Gildas Morvan\footnote{\url{http://www.lgi2a.univ-artois.fr/~morvan/}} during the phase 5 of CISIT\footnote{\url{http://www.cisit.org/}} within the ISART project.

\subsection{Context}

A flow of moving agents can be observed at different scales. Thus, in traffic modeling, three levels are generally considered: the \textit{micro}, \textit{meso} and \textit{macro} levels, representing respectively the interactions between vehicles, groups of vehicles sharing common properties (such as a common destination or a common localization) and flows of vehicles. Each approach is useful in a given context: micro and meso models allow to simulate road networks with complex topologies such as urban area, while macro models allow to develop control strategies to prevent congestion in highways.

\begin{figure}[h]
\begin{center}
\begin{tikzpicture}[>=stealth',shorten >=1pt,auto,semithick]
\node (micro) {\textit{micro}} ;
\node[above of=micro,node distance=1.5cm]  (meso) {\textit{meso}} ;
\node[above of=meso,node distance=1.5cm]   (macro) {\textit{macro}} ;

\node[right of=micro,node distance=3cm]  (vehicle) {vehicle} ;
\node[right of=meso,node distance=3cm]  (group) {group of vehicles} ;
\node[right of=macro,node distance=3cm]  (flow) {flow of vehicles} ;
\node[right of=micro,node distance=8cm]  (instrumentation) {vehicle instrumentation} ;
\node[right of=meso,node distance=8cm]  (panels) {variable-message panels} ;
\node[right of=macro,node distance=8cm]  (metering) {ramp metering} ;

\node[above of=flow,node distance=1.5cm] {\textit{simulated entities}} ;
\node[above of=metering,node distance=1.5cm] {\textit{control strategies}} ;

\draw[<->] (vehicle) -- node {} (group)   ;
\draw[<->] (group) -- node {} (flow)   ;
\draw[<->] (vehicle) -- node {} (instrumentation)   ;
\draw[<->] (group) -- node {} (panels)   ;
\draw[<->] (flow) -- node {} (metering)   ;
\end{tikzpicture} 
\caption{Hybrid traffic simulation and control approach}
\label{mlmtraffic}
\end{center}
\end{figure}
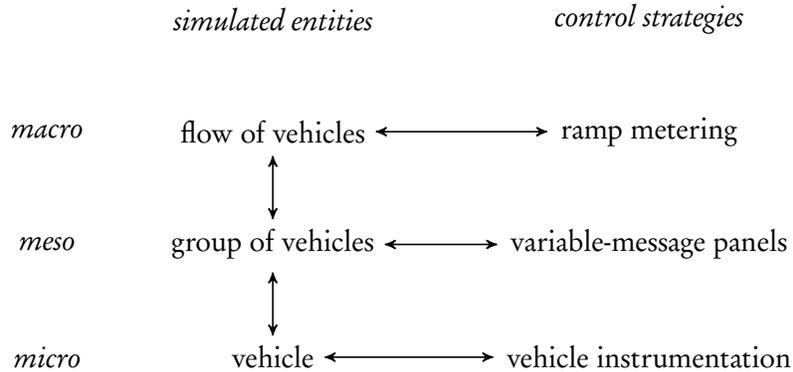

However, to simulate large-scale road networks, it can be interesting to integrate different representations, \textit{e.g.}, micro and macro, in a single model as shown on fig.~\ref{mlmtraffic}. Some existing hybrid micro-macro traffic models  are shown in table~\ref{hybridmodels}.

\begin{table}[h]
\begin{center}
\begin{tabular}{ccc}
\hline
model&micro model&macro model\\
\hline
\citet{Magne:2000}&SITRA-B+&SIMRES\\
\citet{Poschinger:2002}&IDM&Payne\\
\citet{Bourrel:2003}&\multirow{2}{*}{optimal velocity}&LWR\\
\citet{Mammar:2006}&&ARZ\\
\citet{Espie:2006}&ARCHISM&SSMT\\
\citet{El-hmam:2006c}&generic ABM&LWR, ARZ, Payne\\
\hline
\end{tabular}
\caption{Main micro-macro traffic flow models, adapted from~\citet[p. 42]{El-hmam:2006c}} \label{hybridmodels}
\end{center}
\end{table}%

\subsection{Motivations}

The models presented in table~\ref{hybridmodels} share the same limitation: connections between levels are fixed \textit{a priori} and cannot be changed at runtime. Therefore, to be able to observe some emerging phenomena such as congestion formation or to find the exact location of a jam in a large macro section, a dynamic hybrid modeling approach is needed~\citep{Sewall:2011}.
 %(see fig.~\ref{dynamichybrid})
 Multi-level agent-based modeling is an interesting approach to simulate such systems~\citep{Gaud:2008,Gil-Quijano:2010,Gil-Quijano:2012,Picault:2011,Vo:2012,Vo:2012a,Vo:2012b}. Indeed it offers a large range of techniques to dynamically adapt  the level of detail of simulations, couple heterogenous models or detect and reify emergent phenomena~\citep{Morvan:2013}.

Thus, in 2013 we started the development of a multi-level agent-based simulator called JAM-FREE within the ISART project. It allows to simulate large road networks efficiently using a dynamic level of detail.

This simulator relies on a multi-level agent-based modeling framework called SIMILAR --- formely IRM4MLS~\citep{Morvan:2011,Morvan:2012b,Soyez:2013} --- for \textbf{SI}mulations with \textbf{M}ult\textbf{I}-\textbf{L}evel \textbf{A}gents and \textbf{R}eactions. 

% \begin{figure}[H]
%	\begin{center}
%		\includegraphics[height=0.5\textwidth, angle=90]{figures/trafficJamLocationInMacroRepresentation}
%		%\includegraphics[width=\textheight, angle=90]{figures/trafficJamOriginInMicroRepresentation}
%		\caption{A dynamic hybrid modeling use case: follow congestion formation}
%		%\caption{Bibliographical statistics on ML-ABM computed from (a) author's bibliographic database and (b) google scholar data on the January 24, 2013}
%		\label{dynamichybrid}
%	\end{center}
%\end{figure}

The following sections present our two deliverables: SIMILAR and  JAM-FREE.

\section{SIMILAR}

\subsection{Motivations}

The simulation of complex system often requires knowledge coming from 
different sources to obtain relevant results. These sources can either be persons
from different application fields, or different viewpoints on the same phenomenon.

Yet, regular multi-agent based simulation meta models lack the structure to manage 
the aggregation of such systems: their representation of the agents, the 
environment and the temporal dynamics of the system is designed to support a single 
viewpoint.

\begin{figure}[h]
\begin{center}
\includegraphics[width=0.6\textwidth]{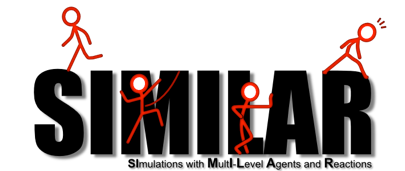}
\caption{SIMILAR logo}
\label{similarlogo}
\end{center}
\end{figure}

To deal with this issue, we developed a generic approach called SIMILAR to design simulations~(fig.~\ref{similarlogo}). This approach relies on a a multi-level, influence-reaction 
and agent-based knowledge representation, more fitting to multiple viewpoints~\citep{Ferber:1996,Michel:2007,Michel:2007a,Morvan:2011,Morvan:2012b,Soyez:2013}.
The approach includes a generic and modular formal model, a methodology and a 
simulation API preserving the structure of the formal model~\citep{Morvan:2014,Morvan:2014a}. Owing to these properties,
the design of the above-mentioned simulations is supported during the whole simulation process,
is more robust to model revisions and relies on a structure fit to represent the intrinsic
complexity of the simulated phenomena.

\subsection{Related works}

Many meta-models and simulation engines dedicated to multi-level agent-based modeling have been proposed in the literature. We reviewed them in~\citet{Morvan:2013}. One can note that these works generally impose constraints on multi-level models, \textit{e.g.}, on interactions or influence graph structure, while SIMILAR allows to model and simulate any multi-level situation as it relies on a formal interaction model~\citep{Ferber:1996}.

\subsection{SIMILAR micro kernel}

SIMILAR is based on the concept of micro kernel. The micro kernel of SIMILAR defines the core classes of the SIMILAR API. It provides a direct correspondence between the concepts developed in the SIMILAR theory and the concrete implementation of simulation. It provides only the bare minimal implementation of these concepts, to leave the implementation of simulation opened to many optimizations. Thus, this kernel is the most appropriate for developers wishing to tune precisely the low level implementation of the simulation.

\begin{figure}[ht]
	\begin{center}
		\includegraphics[width=0.9\textwidth]{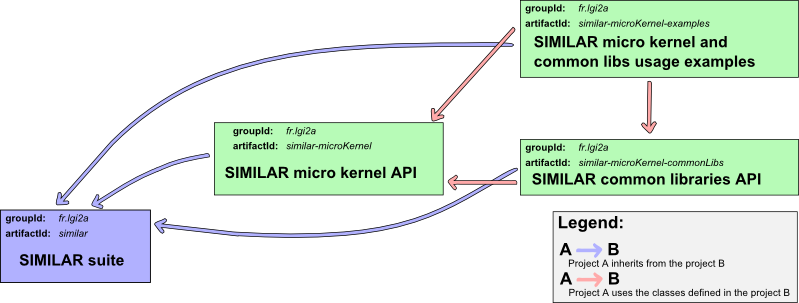}
		\caption{SIMILAR micro kernel structure}
		\label{similarmk}
	\end{center}
\end{figure}

The micro kernel of SIMILAR is characterized by three components (fig.~\ref{similarmk}):
\begin{itemize}
\item The API of the micro kernel, containing the java classes of that kernel.
\item The common libraries of the minimal kernel, containing generic implementations of simulation algorithms and results exporting features.
\item The examples illustrating the use of the micro kernel and the common libraries to design simulations.
\end{itemize}

\subsection{SIMILAR architecture}

\begin{figure}[ht]
	\begin{center}
		\includegraphics[width=0.9\textwidth]{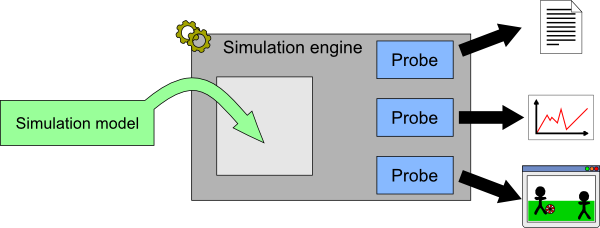}
		\caption{SIMILAR architecture}
		\label{similarclasses}
	\end{center}
\end{figure}

SIMILAR is designed to have a highly customizable architecture, while making the structure of a simulation as explicit as possible. It distinguishes explicitly elements that are often left to the developers and embedded into the code of the simulations:

The concepts of SIMILAR are embodied as concrete classes (fig.~\ref{similarclasses}): 
\begin{itemize}
\item The simulation model, containing the declarative part of the simulation (the description of the simulated phenomenon);
\item The simulation engine, containing the procedural part of the simulation (the generic algorithms running simulations);
\item The observation probes, reading information about the simulation and exporting them in various formats.
\end{itemize}

\subsection{Simulation engine}

In SIMILAR, the notion of simulation model is separated from the notion of simulation engine to separate the declarative knowledge of the simulation from the procedural knowledge of the simulation.

\begin{itemize}
\item A simulation model assembles the simulation-case specific knowledge. It contains the definition of the levels, the agents, the natural action of the environment and the reaction of each level. This knowledge is domain specific.
\item  A simulation engine assembles the execution-related information of a simulation. It defines how time moves, when to ask the agents to perceive, memorize or decide, when the environment produces its natural action or when the reaction is performed. It is also responsible for the observation and the exporting of the simulation data and responsible for the reaction to the so called system influences.
\end{itemize}

The separation between model and engine facilitates the optimization of simulations, by using the most appropriate execution mode depending on the simulation: a simulation engine can be implemented using different inner mechanisms to manage the execution of a simulation.

The simulation engine has different roles in the simulation:

\begin{itemize}
\item Move the simulation through time and call when it is appropriate:
\begin{itemize}
\item The perception, memorization, decision phases of the agents;
\item  The natural phase of the environment;
\item  The reaction phase of the levels;
\end{itemize}
\item Ensure that the time-related constraints of the model remain valid during the whole execution of the simulation. \citet{Morvan:2011} describes the time constraints that the perception, memorization, natural, decision and reaction phases of the simulation have to verify;
\item Ensure that the observation probes are updated appropriately;
 \item Provide a reaction to the system influences of any simulation.
\end{itemize}

\begin{figure}[ht]
	\begin{center}
\begin{tikzpicture} 
\umlemptyclass{AbstractSimulationEngine} 
\umlemptyclass[x=-4, y=-2.3]{MultiCoreCPUSimulationEngine} 
\umlemptyclass[x=4, y=-2.3]{GPGPUSimulationEngine} 
\umlinherit[geometry=|-|]{MultiCoreCPUSimulationEngine}{AbstractSimulationEngine} 
\umlinherit[geometry=|-|]{GPGPUSimulationEngine}{AbstractSimulationEngine} 
\end{tikzpicture}
		\caption{Simulation engines in SIMILAR}
		\label{engine}
	\end{center}
\end{figure}
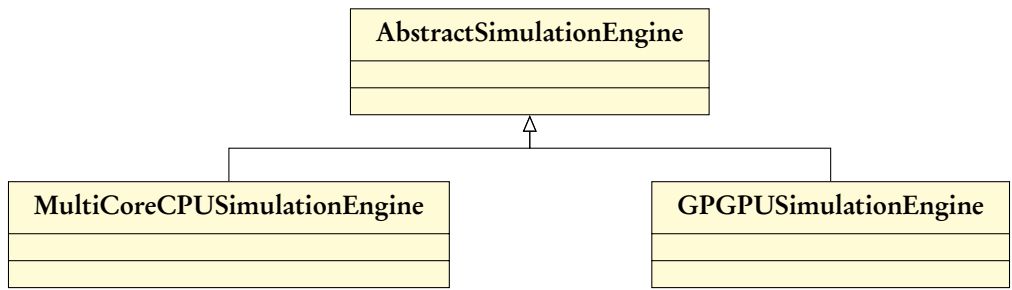

Considering the simulation engine as a top-class abstraction allows to easily simulate a same model on different computing architectures such as multicore CPUs or GPGPU (general purpose graphical processing unit) as shown on fig.~\ref{engine}. 

\subsection{Probes}

The observation of the data of the simulation is managed using probes. Probes are objects listening to the evolution of the state of the simulation. Since the state of the simulation is not always consistent (especially during the computation of the reaction), the probes do not decide by themselves when to observe the simulation. Instead, they are registered to the simulation engine. The simulation engine is responsible to tell probes when the state of the simulation is consistent, and thus when probes can observe, process and export information about the simulation. The moments when the state of the simulation is consistent are:

\begin{itemize}
\item  After the initialization phase of the simulation, but before the execution of the operations related to the first time stamp of the simulation.
\item After the execution of a time stamp of the simulation (i.e. after the computation of the reaction leading to the update of the current time stamp of the simulation)
\item After the execution of the final time stamp of the simulation.
\end{itemize}
Since simulations do not always execute peacefully, probes also have to manage the case when the simulation fails because of an error. Moreover, since some probes might use external resources, they also have to be notified when a new simulation will start. Consequently, probes are also notified when:
\begin{itemize}
\item Each time a new simulation is performed. In reaction, they will be able for instance to open a writing stream to another file, read another configuration file, etc.
\item Each time the simulation stops because of an error. In reaction, they can print the error that caused the simulation to abort, or they can close the streams they were using.
\end{itemize}

\section{JAM-FREE}

In this section we present a multi-level agent-based traffic simulator called JAM-FREE (Java, Agent and Multi-level based Framework for Road-traffic Examination and Enhancement).

\subsection{Motivations}

Our motivation to develop JAM-FREE is to
\begin{enumerate}
\item simulate the road traffic on very large zones including both city and highways while
\item being able to test different signing or traffic redirection strategies,
\item assess their precise influence on the traffic flow and
\item understand why traffic perturbations do occur.
\end{enumerate}

The first goal is easily achieved using macroscopic simulation models at the expense of the fourth goal. Conversely, the fourth goal is easily achieved using microscopic simulation models at the expense of the first goal. To deal with this paradox, we designed an dynamic hybrid model where we benefit from the both approaches.

The advantages of this hybrid approach include the ability:
\begin{itemize}
\item To obtain both quantitative and qualitative information about the road traffic, using respectively macroscopic and microscopic simulation representations in the same simulation.
\item To switch between these representations locally depending:
\begin{itemize}
\item On the simulation needs. For instance understanding the source of a traffic jam.
\item On the computation constraints. For instance managing the CPU load.
\end{itemize}
\item To experiment both macroscopic and microscopic routing strategies, i.e. road load balancing strategies.
\end{itemize}

\subsection{Related works}

To the best of our knowledge, the only other work describing a dynamic hybrid model is~\citep{Sewall:2011}\footnote{Authors maintain a webpage dedicated to this paper: \url{http://gamma.cs.unc.edu/HYBRID_TRAFFIC/}}. However, authors focus on the visualization of data rather than accurate simulation of traffic flow.

\subsection{JAM-FREE use cases}

In this section we present some of the JAM-FREE use cases.

The hybrid model can change the traffic representation of a portion of the road network dynamically. This feature can be used to switch from:
\begin{itemize}
\item a microscopic representation to a macroscopic representation when the CPU is overused in order to reduce its load;
\item a macroscopic representation to a microscopic representation to find out the reason why the traffic jam appeared.
\end{itemize}
These features can be used jointly to balance the load between two clusters.

The clusters used in the hybrid model are not static. Their number can be increased by dividing existing clusters into two or more clusters. This feature can be used to locate traffic jams. Indeed, macroscopic representations provide quantitative information concerning the traffic, including the mean speed and the free driving speed. A significant difference between these values indicates the presence of a traffic jam. Finding the area where the traffic jam is located consists in dividing the clusters and focusing on the ones where the mean speed is significantly lower than the free driving speed.

\subsection{Road network structure}

\begin{figure}[htp]
	\begin{center}
		\includegraphics[width=\textwidth]{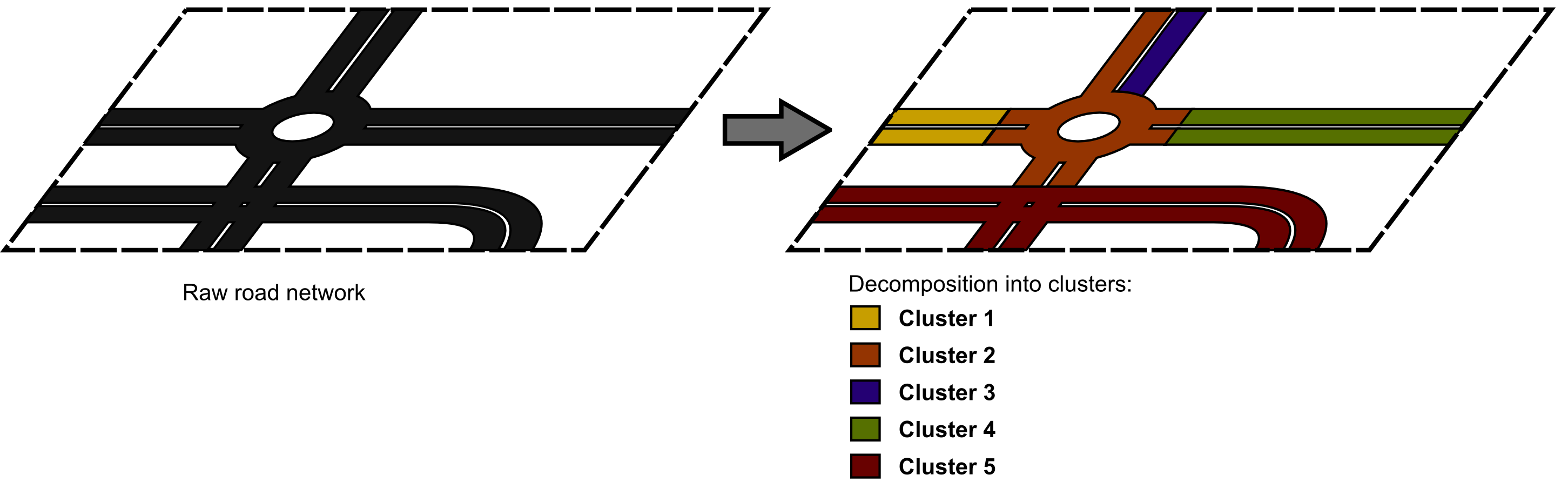}
		\caption{SIMILAR road network structure}
		\label{roadNetworkStructure}
	\end{center}
\end{figure}

\begin{figure}[htp]
	\begin{center}
		\includegraphics[width=0.95\textheight, angle=90]{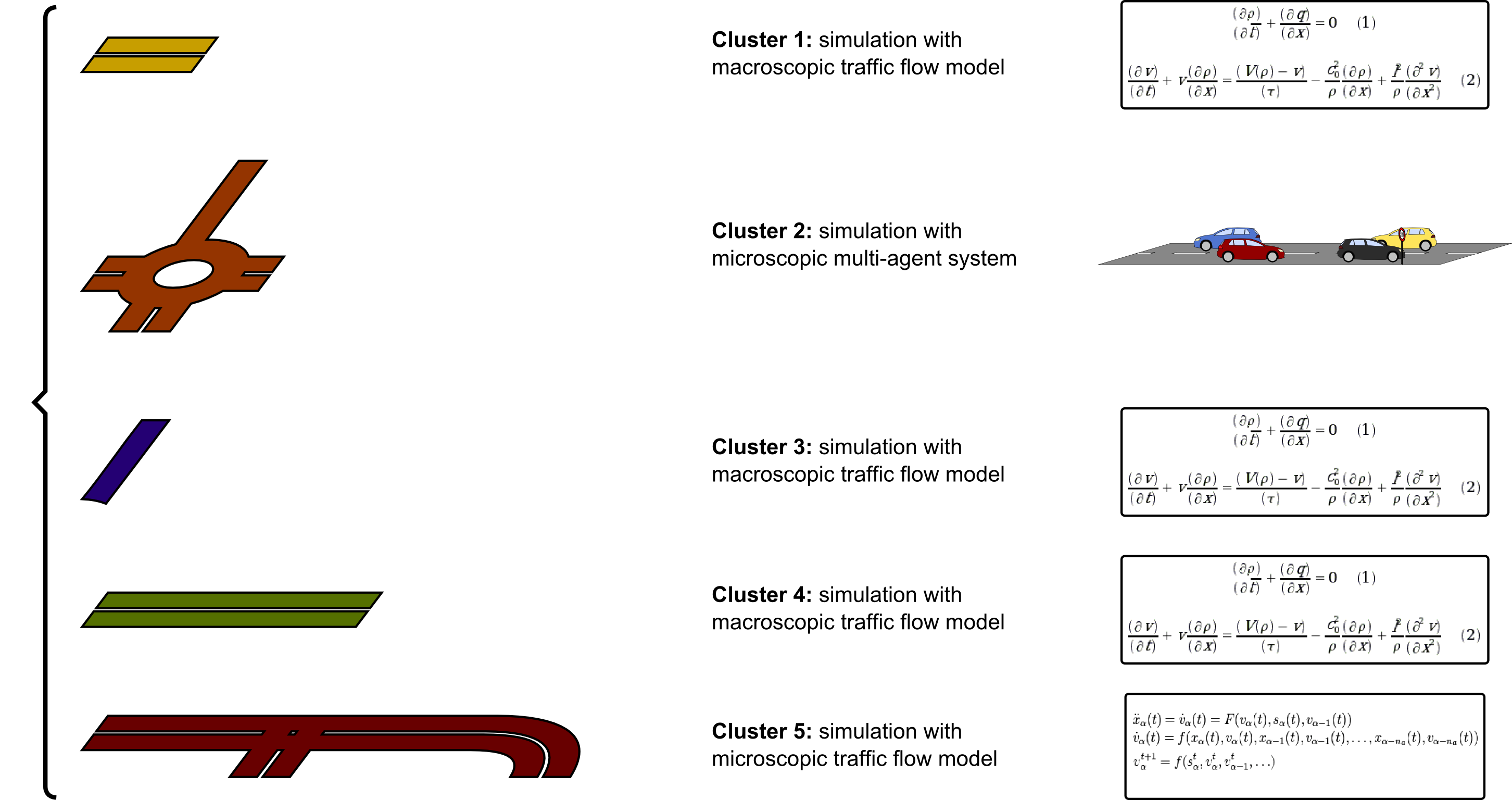}
		\caption{SIMILAR traffic representations}
		\label{trafficRepresentations}
	\end{center}
\end{figure}

We consider that the road network can be divided dynamically into subsets called clusters (see fig.~\ref{roadNetworkStructure}). The traffic is simulated on each cluster using various heterogeneous models depending on the situation: either a microscopic model or a macroscopic model (see fig.~\ref{trafficRepresentations}).

For computation time efficiency reasons, we choose not to model the road network at a physical level. Indeed, such a model requires the vehicles to interpret a large continuous zone of asphalt with blank lines as separated lanes. Such computations are unnecessary complex, since in our use case (France) rational drivers usually never drive over the blank lines separating the lanes if they are not overtaking. To keep the model simple, the road network is only modeled at a semantic level.

The road network is modeled as a set of interconnected roads. Each road contains a set of lanes (usually from 1 to 5). Roads also contain vertical signs (e.g. stop sign, speed limit sign) that are bond either to all the lanes or to specific lanes (e.g. extraction lane in the highway, slow vehicles lane). In our model, we support a specific subset of the french vertical signs .
The connection points of the roads are called nodes. We distinguish different types of nodes:
\begin{itemize}
\item Crossroads nodes
\item Roundabout nodes
\item Highway insertion node
\item Highway extraction node
\end{itemize}

The naming and identification of road network components follow the Sétra\footnote{Sétra (Service d'études sur les transports, les routes et leurs aménagements) is a technical service of the  Ministère de l'Écologie, du Développement Durable et de l'Énergie dedicated to transportation issues: \url{http://www.setra.developpement-durable.gouv.fr}} specification\footnote{\url{http://dtrf.setra.fr/pdf/pj/Dtrf/0005/Dtrf-0005792/DT5792.pdf}}~\citep{Setra:2010}.

\subsection{Vehicle behaviors}

\begin{figure}[htp]
	\begin{center}
		\includegraphics[width=\textwidth]{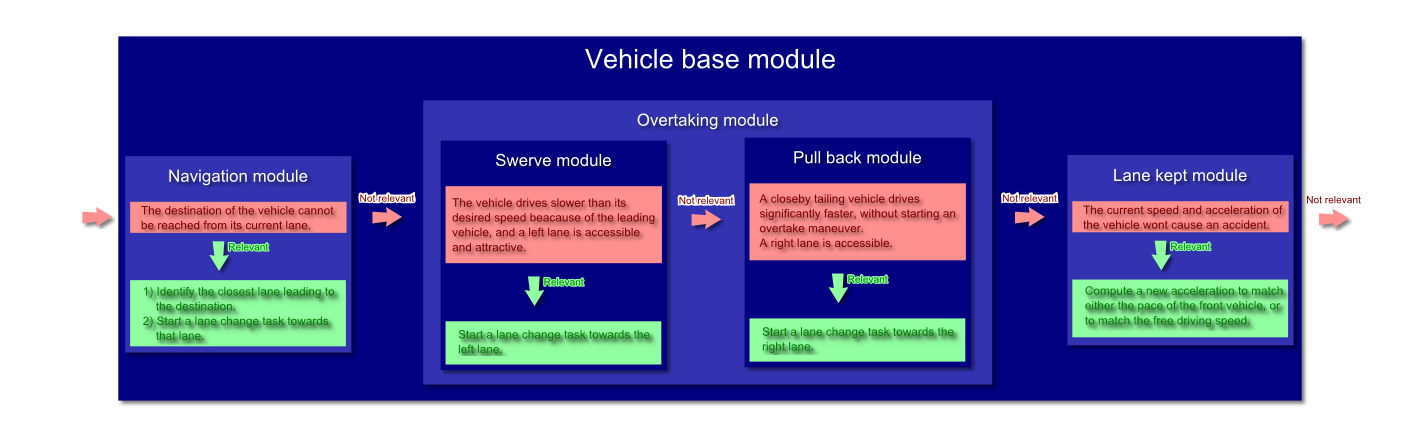}
		\caption{Vehicle behavior structure in SIMILAR}
		\label{behaviorStructure}
	\end{center}
\end{figure}

The behavior of a vehicle in our simulation is designed to be modular, customizable and easily extendible. To achieve these properties, the behavior of a vehicle relies on the following "chain of responsibility" design pattern (see fig.~\ref{behaviorStructure}).

The behavior of a vehicle can be seen as the sum of three different parts:
\begin{itemize}
\item A navigation behavior when the current lane of the vehicle does not lead to the desired destination of the vehicle. This behavior consists in changing the lane of the vehicle until the current lane of the vehicle leads to the desired destination.
\item An overtaking behavior when the vehicle either wants to increase its speed by changing its lane, or when the vehicle has to swerve because a faster vehicle is tailing it.
\item A behavior when no lane change is required (acceleration model).
\end{itemize}

\subsubsection{Acceleration model}

JAM-FREE implements the highly used Intelligent-Driver Model (IDM) to model the acceleration behavior of vehicles~\citep{Kesting:2010}.

The IDM is a microscopic traffic flow model, i.e., each vehicle-driver combination constitutes an active "particle" in the simulation. Such model characterize the traffic state at any given time by the positions and speeds of all simulated vehicles. In case of multi-lane traffic, the lane index complements the state description. More specifically, the IDM is a car-following model. In such models, the decision of any driver to accelerate or to brake depends only on his or her own speed, and on the position and speed of the "leading vehicle" immediately ahead. Lane-changing decisions, however, depend on all neighboring vehicles (see the lane-changing model MOBIL). The model structure of the IDM can be described as follows:

\begin{itemize}
\item The influencing factors (model input) are the own speed v, the bumper-to-bumper gap s to the leading vehicle, and the relative speed (speed difference) of the two vehicles (positive when approaching).
\item The model output is the acceleration chosen by the driver for this situation.
\item The model parameters describe the driving style, i.e., whether the simulated driver drives slow or fast, careful or reckless
\end{itemize}

\subsubsection{Lane changing model}

Lane changes takes place, if:
\begin{itemize}
\item the potential new target lane is more attractive, i.e., the "incentive criterion" is satisfied,
\item and the change can be performed safely, i.e., the "safety criterion" is satisfied.
\end{itemize}

JAM-FREE implements a modified version of the lane changing model MOBIL\footnote{\url{http://xxx.uni-augsburg.de/abs/cond-mat/0002177}}.

Chosen acceleration and lane changing models are presented in detail in the JAM-FREE documentation.

\subsection{Traffic generation}

In our model, each Traffic start connector (see this page) is a special connector that will generate traffic on the road it is attached to. Note that such a connector has to be created and put on each lane where the traffic appears. The creation of vehicles is managed by a traffic input point agent. 

Various implementation of this agent currently exist:
\begin{itemize}
\item "Flow-mass traffic input point " agents, generating the traffic flow using a flow-mass parameter
\item "Scripted traffic input point " agents, generating the traffic flow using user-defined events. The created vehicles and the creation dates are manually specified by the users.
\end{itemize}

\subsection{User interfaces}

JAM-FREE simulations and models can be created using a convenient hybrid XML/Java system:
\begin{itemize}
\item simple simulations can be created using only XML files,
\item complex simulations (for instance with dedicated vehicle generation methods, or behaviors) can be created by implementing Java classes that are referred in the XML files.
\end{itemize}

The XML files describe the "physical" road network as well as the simulation itself (traffic generation methods and models). An example of simulation is shown in appendix~\ref{app:xml}.

To simply manage simulations, a graphical user interface called JFF (JAM-FREE Framework) has been developed (see appendix~\ref{app:jff}).

\section{Conclusion}

Within the ISART project of CISIT, we developed a multi-level agent-based traffic simulator called JAM-FREE, relying on a dedicated meta-model called SIMILAR. These software programs have been developed using state of the art software engineering tools such as SonarQube\footnote{\url{http://www.sonarqube.org}} (see appendix~\ref{app:SonarQube}), Maven\footnote{\url{http://maven.apache.org}}, Jenkins\footnote{\url{jenkins-ci.org}} or Hamcrest\footnote{\url{https://code.google.com/p/hamcrest/}}. Massive unit tests were carried to verify simulation engines and models. For instance, 545800 tests (265Mo of source code), each describing a specific use case, have been generated to validate the perception model of JAM-FREE.

\subsection{Deliverables}

The meta-model and simulation engine of SIMILAR have been implemented in Java, and will be available freely in a few weeks under the CeCILL-B\footnote{\url{ http://cecill.info/licences/Licence_CeCILL-B_V1-en.html}} license. The project is currently hosted on \url{https://forge.univ-artois.fr} with a limited access to the source code.

JAM-FREE is still under development. However, a preliminary demo version can be distributed to the CISIT partners if needed.

Two papers are in preparation: 
\begin{itemize}
\item \citet{Morvan:2014}, that describes the SIMILAR meta-model,
\item \citet{Morvan:2014a}, that proposes novel ways to manage uncertainty in multi-agent-based simulations.
\end{itemize}

\subsection{Perspectives}

During the next phase of CISIT, we will pursue the development of JAM-FREE. For instance, one of our goal is to be able to import data from the open source geographical information system OpenStreetMap\footnote{\url{http://www.openstreetmap.org}} and directly use it in JAM-FREE.

\begin{appendices}

\section{JAM-FREE XML files}
\label{app:xml}
A minimal JAM-FREE simulation is defined by a set of XML files. A simple example is given in the following. The main file specifies simulation's main options as well as the XML files describing the road network structure and traffic generation.

\begin{lstlisting}[caption=JAM-FREE: main XML file]
<?xml version="1.0" encoding="UTF-8"?>
<jamFreeModel>

	<!-- Identify the file where the infrastructure of the road network is described. -->
	<roadNetwork fileName="infrastructure/navigation-infrastructure.xml" />
	
	<!-- Describe how time moves through simulation -->
	<time sampling="0.05"> <!-- A time step occurs every 0.05 second of the simulated world (optional). -->
		<simulationEnd time="180" /> <!-- The simulation runs during 60 seconds. -->
		<microscopicTrafficEvolution period="1" /> <!-- The "microscopic - traffic" level acts at each time step (optional) -->
	</time>
	
	<!-- Define the initial state of the different levels of the simulation. -->
	<levels>
		<!-- Identify the file where the initial state of the "microscopic - traffic" level is described. -->
		<microscopicTrafficLevel fileName="microscopicLevel/microscopicLevel-example1.xml" />
	</levels>
	
	<!-- Define how the results of the simulation can be seen. -->
	<probes />
	
</jamFreeModel>
\end{lstlisting}

The infrastructure files describes the road components contained in the road network and their connections.

\begin{lstlisting}[caption=JAM-FREE: infrastructure XML file]
<?xml version="1.0" encoding="UTF-8"?>
<roadnetwork>
<!-- 
	Description of the road components contained in the road network. 
-->
	<road id="N9PR1_section1" numberOfLanes="2" width="4" semanticLength="1000">
		<type value="national"/>
		<input plo="N9PR1" distanceFromPlo="0">
			<laneIdRange from="0" to="1"/>
		</input>
		<output plo="N9PR1" distanceFromPlo="1000">
			<laneIdRange from="0" to="1"/>
		</output>
		<shape>
			<point x="0" y="8"/>
			<point x="1000" y="8"/>
		</shape>
	</road>
	<road id="N9PR1_section2" numberOfLanes="2" width="4" semanticLength="500">
		<type value="national"/>
		<input plo="N9PR1" distanceFromPlo="1000">
			<laneIdRange from="0" to="1"/>
		</input>
		<output plo="N9PR1" distanceFromPlo="1500">
			<laneIdRange from="0" to="1"/>
		</output>
		<reachableDestinations>
			<laneDestination>
				<laneId id="1" />
				<destination name="Paris" />
			</laneDestination>
			<laneDestination>
				<laneId id="0" />
				<destination name="Dunkerque" />
				<destination name="Arras" />
			</laneDestination>
		</reachableDestinations>
		<shape>
			<point x="1000" y="8"/>
			<point x="1500" y="8"/>
		</shape>
	</road>
	<road id="D19PR1_section1" numberOfLanes="1" width="4" semanticLength="300">
		<type value="national"/>
		<input plo="N9PR1" distanceFromPlo="1500">
			<laneId id="1"/>
		</input>
		<output plo="D19PR1" distanceFromPlo="300">
			<laneId id="0"/>
		</output>
		<reachableDestinations>
			<laneDestination>
				<laneId id="1" />
				<destination name="Paris" />
			</laneDestination>
		</reachableDestinations>
		<shape>
			<point x="1500" y="4"/>
			<point x="1800" y="4"/>
		</shape>
	</road>
	<road id="D20PR1_section1" numberOfLanes="1" width="4" semanticLength="300">
		<type value="national"/>
		<input plo="N9PR1" distanceFromPlo="1500">
			<laneId id="0"/>
		</input>
		<output plo="D20PR1" distanceFromPlo="300">
			<laneId id="0"/>
		</output>
		<reachableDestinations>
			<laneDestination>
				<laneId id="0" />
				<destination name="Dunkerque" />
				<destination name="Arras" />
			</laneDestination>
		</reachableDestinations>
		<shape>
			<point x="1500" y="8"/>
			<point x="1501" y="8"/>
			<point x="1651" y="158"/>
		</shape>
	</road>

<!-- 
	Description of the connection between the road components.
-->
	<connection outputComponentId="N9PR1_section1" inputComponentId="N9PR1_section2">
		<input plo="N9PR1" distanceFromPlo="1000">
			<laneIdRange from="0" to="1"/>
		</input>
		<output plo="N9PR1" distanceFromPlo="1000">
			<laneIdRange from="0" to="1"/>
		</output>
	</connection>
	<connection outputComponentId="N9PR1_section2" inputComponentId="D19PR1_section1">
		<input plo="N9PR1" distanceFromPlo="1500">
			<laneId id="1"/>
		</input>
		<output plo="N9PR1" distanceFromPlo="1500">
			<laneId id="1"/>
		</output>
	</connection>
	<connection outputComponentId="N9PR1_section2" inputComponentId="D20PR1_section1">
		<input plo="N9PR1" distanceFromPlo="1500">
			<laneId id="0"/>
		</input>
		<output plo="N9PR1" distanceFromPlo="1500">
			<laneId id="0"/>
		</output>
	</connection>

	<road id="D19PR1_section2" numberOfLanes="2" width="4" semanticLength="500">
		<type value="national"/>
		<input plo="D19PR1" distanceFromPlo="300">
			<laneIdRange from="0" to="1" />
		</input>
		<output plo="D19PR1" distanceFromPlo="800">
			<laneIdRange from="0" to="1" />
		</output>
		<explicitSpecification>
			<!-- We tell that the lane having input point "0" starts at the beginning of the road. -->
			<laneStartingPoint inputId="0" />
		</explicitSpecification>
		<shape>
			<point x="1800" y="8"/>
			<point x="2300" y="8"/>
		</shape>
	</road>
	<road id="D19PR1_section3" numberOfLanes="2" width="4" semanticLength="500">
		<type value="national"/>
		<input plo="D19PR1" distanceFromPlo="800">
			<laneIdRange from="0" to="1" />
		</input>
		<output plo="D19PR1" distanceFromPlo="1300">
			<laneIdRange from="0" to="1" />
		</output>
		<reachableDestinations>
			<laneDestination>
				<laneId id="1" />
				<destination name="Paris" />
			</laneDestination>
			<laneDestination>
				<laneId id="0" />
				<destination name="Lille" />
			</laneDestination>
		</reachableDestinations>
		<shape>
			<point x="2300" y="8"/>
			<point x="2800" y="8"/>
		</shape>
	</road>
	<connection outputComponentId="D19PR1_section1" inputComponentId="D19PR1_section2">
		<input plo="D19PR1" distanceFromPlo="300">
			<laneId id="1"/>
		</input>
		<output plo="D19PR1" distanceFromPlo="300">
			<laneId id="0"/>
		</output>
	</connection>
	<connection outputComponentId="D19PR1_section2" inputComponentId="D19PR1_section3">
		<input plo="D19PR1" distanceFromPlo="800">
			<laneIdRange from="0" to="1"/>
		</input>
		<output plo="D19PR1" distanceFromPlo="800">
			<laneIdRange from="0" to="1"/>
		</output>
	</connection>
	
	
	<road id="D25PR1_section1" numberOfLanes="1" width="4" semanticLength="300">
		<type value="national"/>
		<input plo="D19PR1" distanceFromPlo="1300">
			<laneId id="0"/>
		</input>
		<output plo="D25PR1" distanceFromPlo="300">
			<laneId id="0"/>
		</output>
		<reachableDestinations>
			<laneDestination>
				<laneId id="0" />
				<destination name="Lille" />
			</laneDestination>
		</reachableDestinations>
		<shape>
			<point x="2800" y="8"/>
			<point x="2802" y="8"/>
			<point x="2802" y="306"/>
		</shape>
	</road>
	<connection outputComponentId="D19PR1_section3" inputComponentId="D25PR1_section1">
		<input plo="D19PR1" distanceFromPlo="1300">
			<laneId id="0"/>
		</input>
		<output plo="D19PR1" distanceFromPlo="1300">
			<laneId id="0"/>
		</output>
	</connection>
	<road id="D26PR1_section1" numberOfLanes="1" width="4" semanticLength="300">
		<type value="national"/>
		<input plo="D19PR1" distanceFromPlo="1300">
			<laneId id="1"/>
		</input>
		<output plo="D26PR1" distanceFromPlo="300">
			<laneId id="0"/>
		</output>
		<reachableDestinations>
			<laneDestination>
				<laneId id="1" />
				<destination name="Paris" />
			</laneDestination>
		</reachableDestinations>
		<shape>
			<point x="2800" y="4"/>
			<point x="3100" y="4"/>
		</shape>
	</road>
	<connection outputComponentId="D19PR1_section3" inputComponentId="D26PR1_section1">
		<input plo="D19PR1" distanceFromPlo="1300">
			<laneId id="1"/>
		</input>
		<output plo="D19PR1" distanceFromPlo="1300">
			<laneId id="1"/>
		</output>
	</connection>
</roadnetwork>
\end{lstlisting}

The microscopic level files defines vehicle generation and destruction points, \textit{i.e.}, the boundaries of the microscopic representation.

\begin{lstlisting}[caption=JAM-FREE: microscopic level XML file]
<?xml version="1.0" encoding="UTF-8"?>

<microTrafficLevel>
<!-- 
	Create the generation points at the beginning of the simulated A1 highway (at the PLO A1PR35D) 
-->
	<trafficGenerationPoint>
		<location>
			<position roadId="N9PR1_section1" plo="N9PR1" distanceFromPlo="0" laneId="0" />
		</location>
		<behavior>
			<rhythmModel 
				javaClass="fr.lgi2a.jamFree.libraries.microscopicLevel.generationRythmModels.FlowMassBased_RhythmModel" 
				parametersFile="microscopicLevel/generationPoints/A1PR35D_section1-rhythmParameters.xml" 
			/>
			<vehicleGenerationModel 
				javaClass="fr.lgi2a.jamFree.libraries.microscopicLevel.vehicleGenerationModels.UniformWithIDMForwardAcceleration_WithMobilLaneChange_SpecificDestination_GenerationModel" 
				parametersFile="microscopicLevel/generationPoints/A1PR35D_section1-generationParameters.xml" 
			/>
		</behavior>
	</trafficGenerationPoint>

<!-- 
	Create the destruction point at the end of the simulated A1 national road (at the PLO A1PR38D)
 -->
	<trafficDestructionPoint>
		<location>
			<position roadId="D20PR1_section1" plo="D20PR1" distanceFromPlo="300" laneId="0" />
		</location>
	</trafficDestructionPoint>
	<trafficDestructionPoint>
		<location>
			<position roadId="D26PR1_section1" plo="D26PR1" distanceFromPlo="300" laneId="0" />
		</location>
	</trafficDestructionPoint>
	<trafficDestructionPoint>
		<location>
			<position roadId="D25PR1_section1" plo="D25PR1" distanceFromPlo="300" laneId="0" />
		</location>
	</trafficDestructionPoint>
</microTrafficLevel>
\end{lstlisting}

Finally, each generation point is characterized by two XML files specifying how (generation parameters) and when (rhythm parameters) vehicles are created.

\begin{lstlisting}[caption=JAM-FREE: generation parameters XML file]
<?xml version="1.0" encoding="UTF-8" standalone="no"?>
<!DOCTYPE properties SYSTEM "http://java.sun.com/dtd/properties.dtd">
<properties>
   <entry key="headwayTimeUsedInGenerationValidation">1.5</entry>
   <entry key="vehicleMinimalGapUsedInGenerationValidation">2.0</entry>

   <entry key="vehicleLength">4.13</entry>
   <entry key="accelerationModel_initialSpeed">20.0</entry>
   
   <entry key="accelerationModel_maximalAcceleration">1.962</entry>
   <entry key="accelerationModel_maximalDeceleration">5.886</entry>
   <entry key="accelerationModel_maximalSpeed">36.1111</entry>
   <entry key="accelerationModel_minimalVehicleGap">2.0</entry>
   <entry key="accelerationModel_headwayTime">1.5</entry>
   
   <entry key="lanceChange_accelerationAnticipationModel_maximalAcceleration">0.2</entry>
   <entry key="lanceChange_accelerationAnticipationModel_maximalDeceleration">3.0</entry>
   <entry key="lanceChange_accelerationAnticipationModel_headwayTime">1.4</entry>
   <entry key="lanceChange_accelerationAnticipationModel_minimalVehicleGap">2.5</entry>
   
   <entry key="lanceChange_maneuverDuration">2.5</entry>
   <entry key="laneChange_maximumSafeDeceleration">4.0</entry>
   <entry key="laneChange_politenessFactor">0.35</entry>
   <entry key="laneChange_accelerationThreshold">0.2</entry>
   <entry key="laneChange_rightLaneAccelerationBias">0.1</entry>
   
   <entry key="laneChange_thresholdDistanceComputation_headwayTime">1.5</entry>
   <entry key="laneChange_thresholdDistanceComputation_minimalVehicleGap">2.0</entry>
   
   <entry key="checkPoints">Paris</entry>
</properties>
\end{lstlisting}

\begin{lstlisting}[caption=JAM-FREE: generation rhythm XML file]
<?xml version="1.0" encoding="UTF-8" standalone="no"?>
<!DOCTYPE properties SYSTEM "http://java.sun.com/dtd/properties.dtd">
<properties>
   <entry key="flowMass">0.20</entry>
</properties>
\end{lstlisting}

\section{JAM-FREE Framework}
\label{app:jff}

\begin{figure}[htp]
	\begin{center}
		\includegraphics[width=0.95\textheight, angle=90]{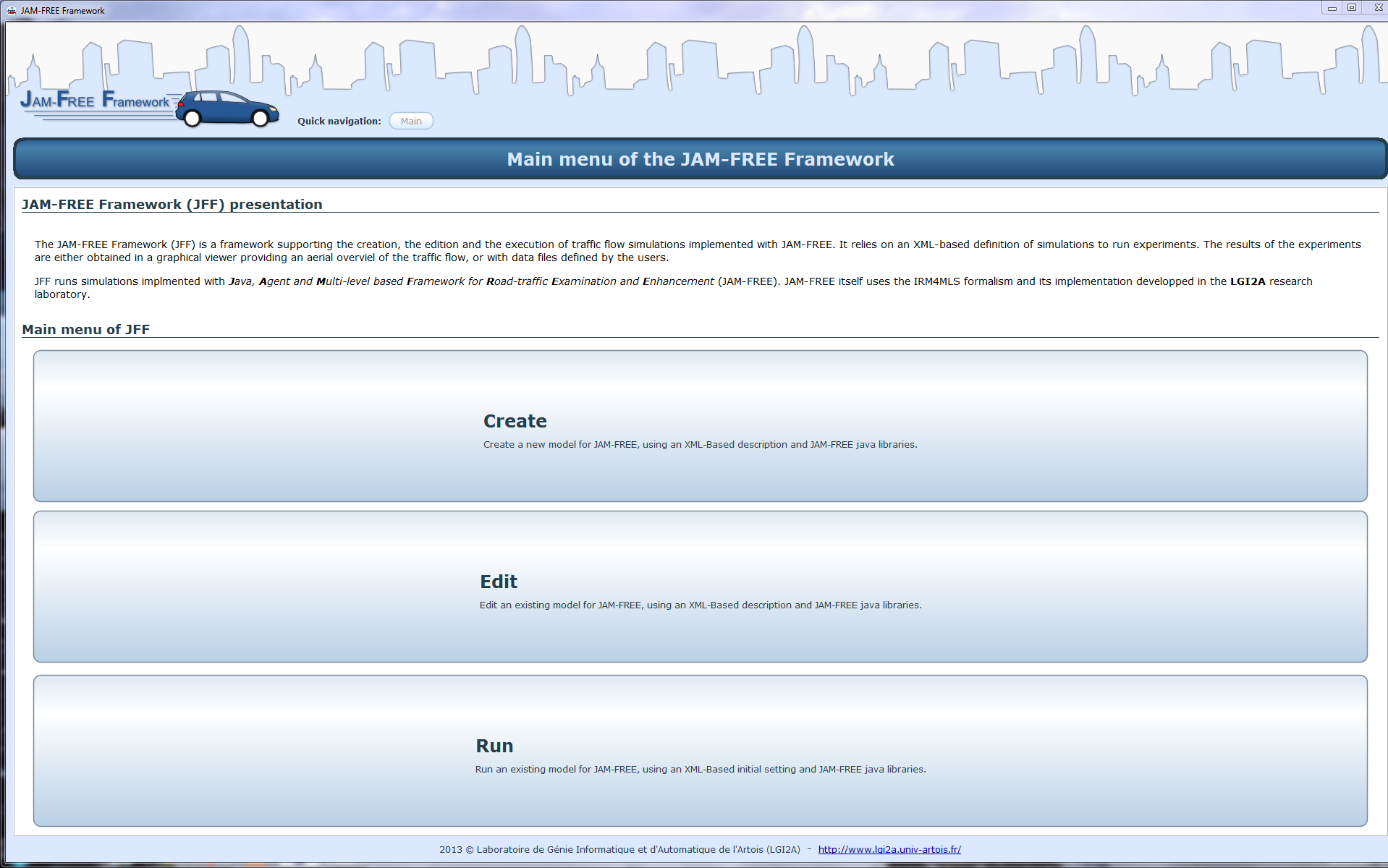}
		\caption{Screenshot of JFF: main menu}
		\label{jffScreenshot_1}
	\end{center}
\end{figure}

\begin{figure}[htp]
	\begin{center}
		\includegraphics[width=0.95\textheight, angle=90]{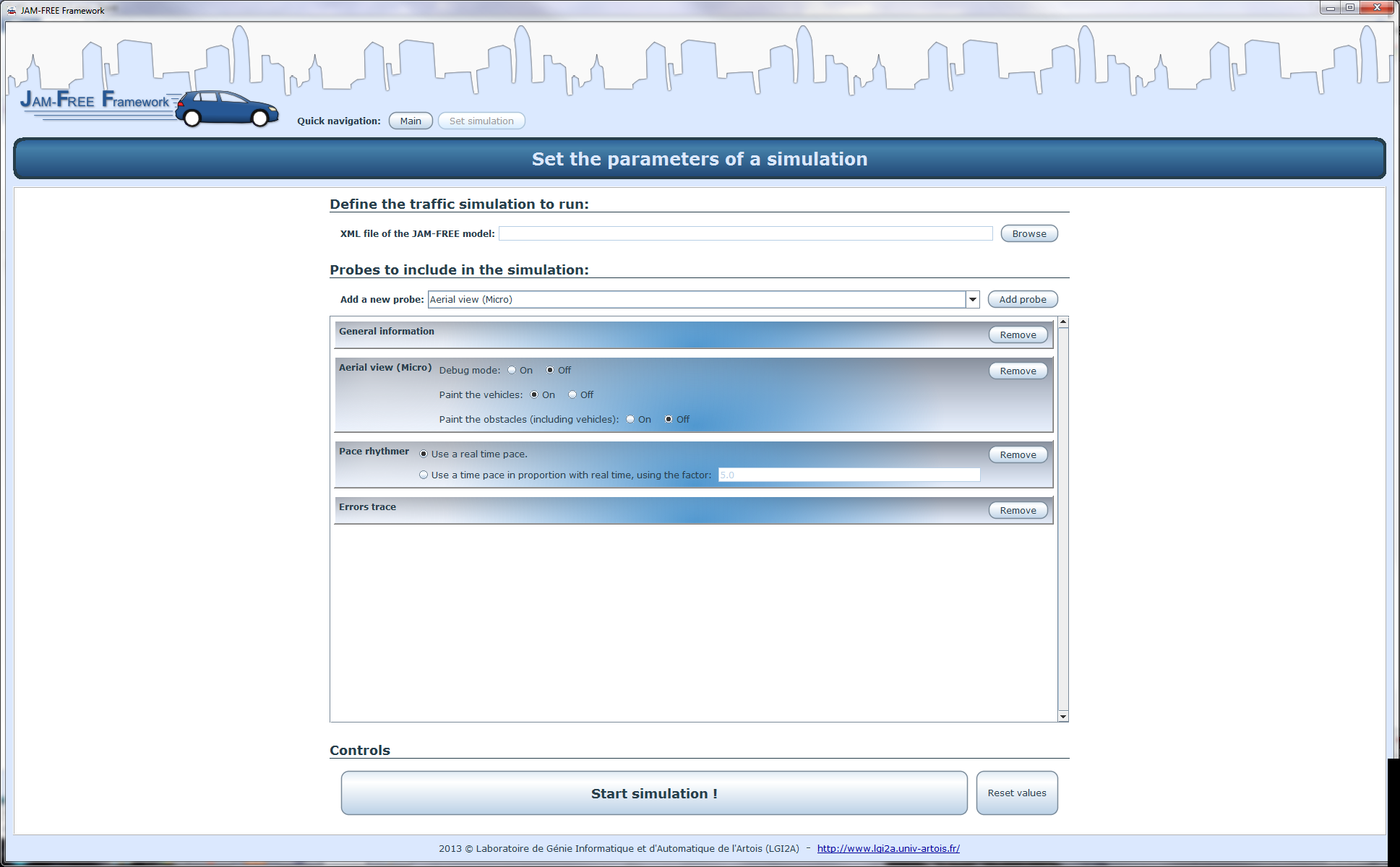}
		\caption{Screenshot of JFF: set the parameters of a simulation}
		\label{jffScreenshot_2}
	\end{center}
\end{figure}

\begin{figure}[htp]
	\begin{center}
		\includegraphics[width=0.95\textheight, angle=90]{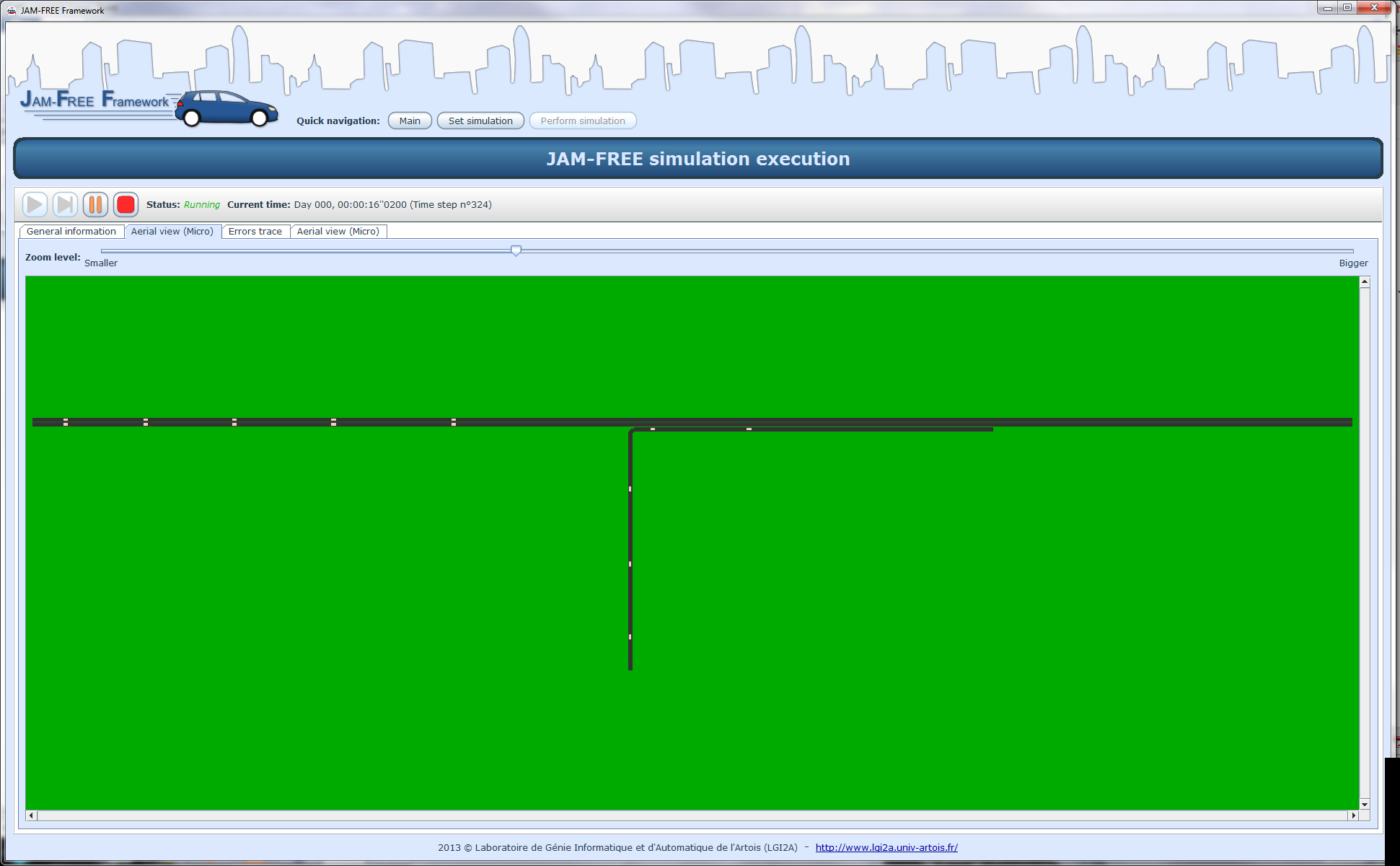}
		\caption{Screenshot of JFF: simulation execution}
		\label{jffScreenshot_3}
	\end{center}
\end{figure}

\begin{figure}[htp]
	\begin{center}
		\includegraphics[width=0.95\textheight, angle=90]{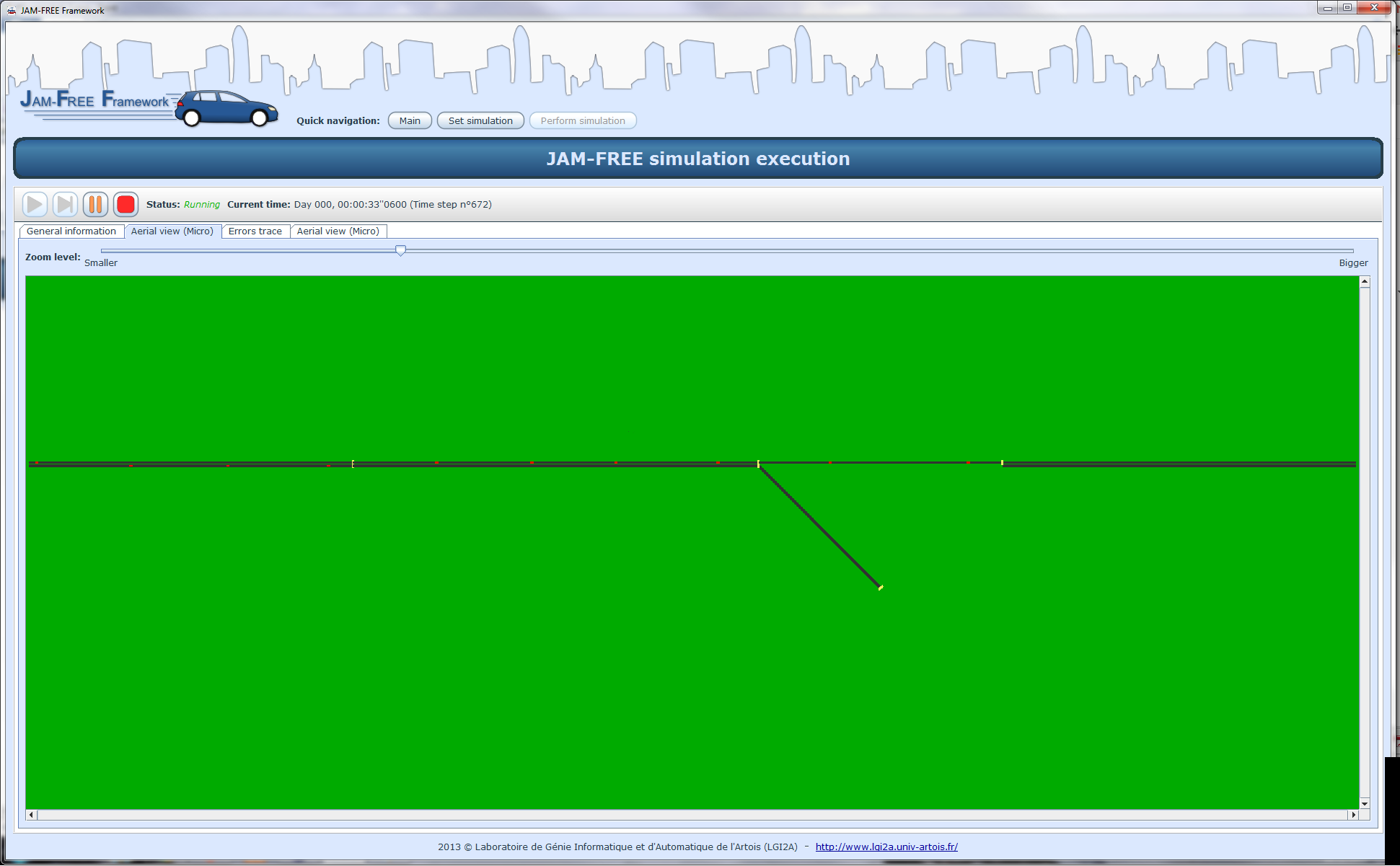}
		\caption{Screenshot of JFF: simulation execution (debug mode)}
		\label{jffScreenshot_4}
	\end{center}
\end{figure}

JAM-FREE Framework (JFF) is a graphical user interface  that allow users to simply manage simulations~(load, run, stop, analyse results). Screenshots of JFF are shown in figures \ref{jffScreenshot_1}--\ref{jffScreenshot_4}.

\section{SonarQube screenshots}
\label{app:SonarQube}

\begin{figure}[htp]
	\begin{center}
		\includegraphics[width=0.95\textheight, angle=90]{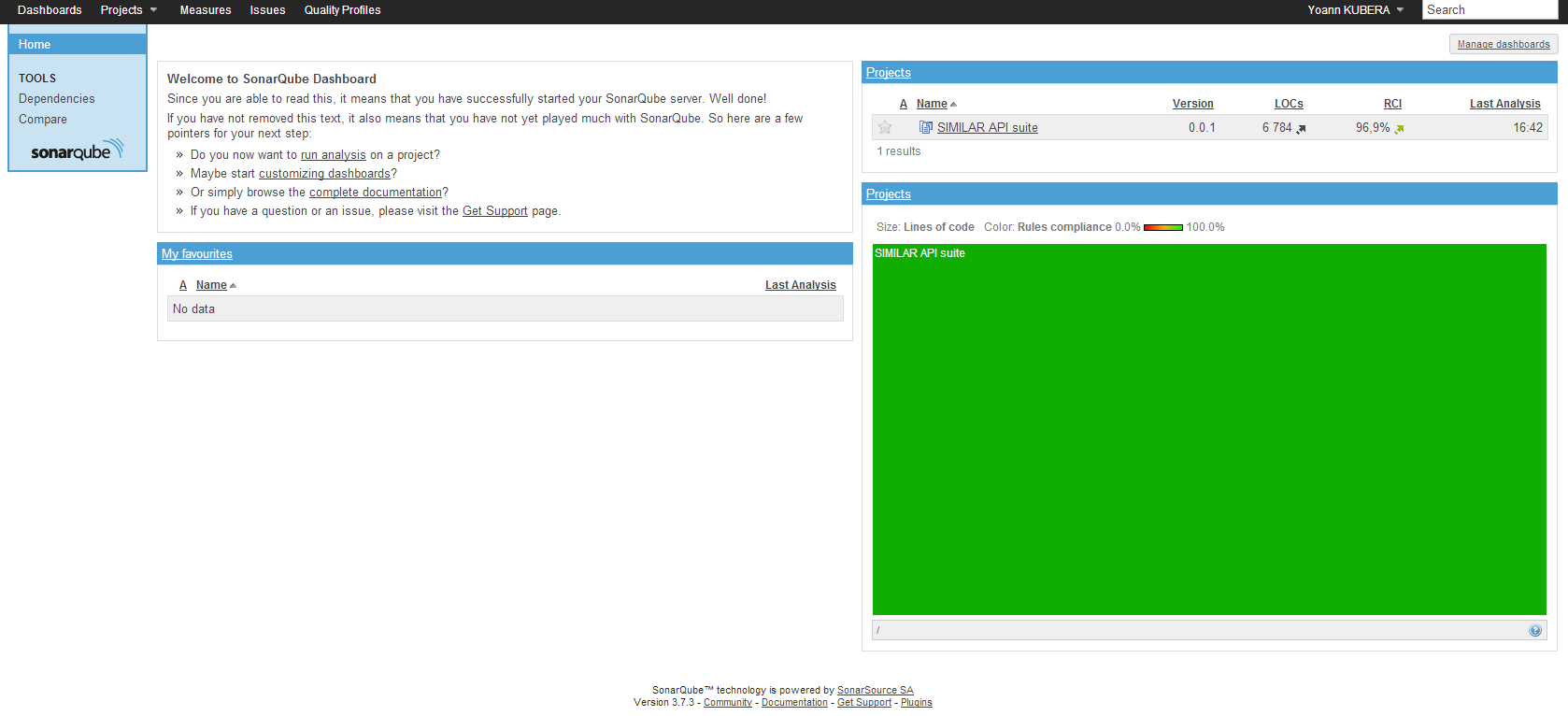}
		\caption{Screenshot of SonarQube: Main screen that shows that SIMILAR compiles at $96.9\%$ to the best coding practices}
		\label{sonar1}
	\end{center}
\end{figure}

\begin{figure}[htp]
	\begin{center}
		\includegraphics[width=0.95\textheight, angle=90]{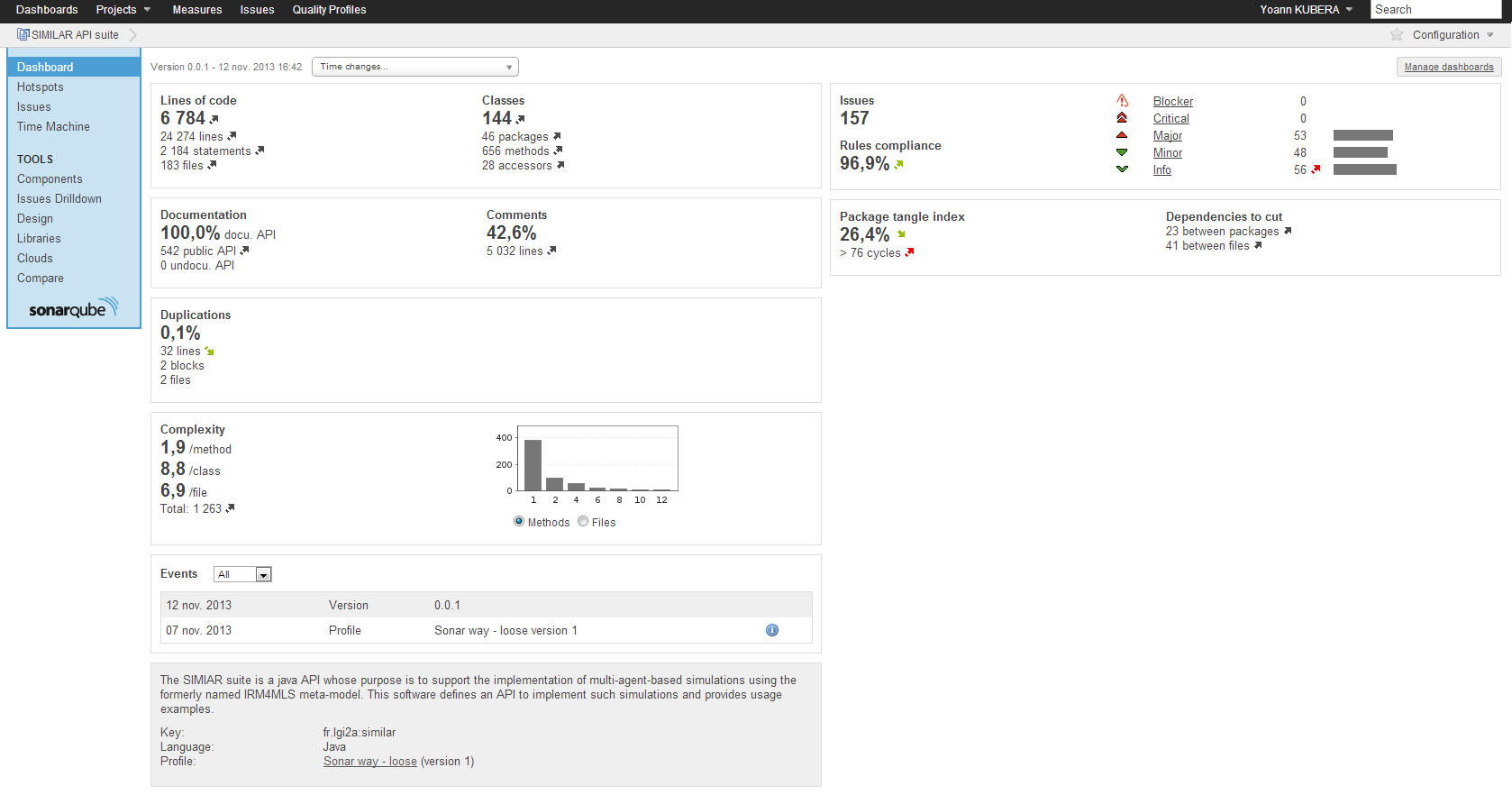}
		\caption{Screenshot of SonarQube: Screen showing the different metrics of SIMILAR}
		\label{sonar2}
	\end{center}
\end{figure}

SonarQube is an open source platform for Continuous Inspection (CI) of code quality. It computes many interesting quality metrics such as the pertcentage of cases covered by unit tests. As the fig.\ref{sonar1} and \ref{sonar2} show, SIMILAR obtains excellent reports on SonarQube.

\end{appendices} 

%\pagebreak
\addcontentsline{toc}{section}{References}
\bibliographystyle{apalike}
%{\scriptsize 
\bibliography{../../Biblio}

\begin{thebibliography}{}

\bibitem[Bourrel and Lesort, 2003]{Bourrel:2003}
Bourrel, E. and Lesort, J. (2003).
\newblock Mixing micro and macro representations of traffic flow: a hybrid
  model based on the {LWR} theory.
\newblock {\em 82th Annual Meeting of the Transportation Research Board}.

\bibitem[{El hmam}, 2006]{El-hmam:2006c}
{El hmam}, M. (2006).
\newblock {\em Contribution {\`a} la mod{\'e}lisation et {\`a} la simulation
  hybride du flux de trafic}.
\newblock PhD thesis, Universit{\'e} d'Artois.

\bibitem[Espi{\'e} et~al., 2006]{Espie:2006}
Espi{\'e}, S., Gattuso, D., and Galante, F. (2006).
\newblock Hybrid traffic model coupling macro- and behavioral microsimulation.
\newblock In {\em 85th Annual Meeting of Transportation Research Board},
  Washington D.C.

\bibitem[Ferber and M\"uller, 1996]{Ferber:1996}
Ferber, J. and M\"uller, J.-P. (1996).
\newblock Influences and reaction: a model of situated multiagent systems.
\newblock In {\em 2nd International Conference on Multi-agent systems
  (ICMAS'96)}, pages 72--79.

\bibitem[Gaud et~al., 2008]{Gaud:2008}
Gaud, N., Galland, S., Gechter, F., Hilaire, V., and Koukam, A. (2008).
\newblock Holonic multilevel simulation of complex systems : Application to
  real-time pedestrians simulation in virtual urban environment.
\newblock {\em Simulation Modelling Practice and Theory}, 16:1659--1676.

\bibitem[{Gil-Quijano} et~al., 2010]{Gil-Quijano:2010}
{Gil-Quijano}, J., Hutzler, G., and Louail, T. (2010).
\newblock Accroche-toi au niveau, j'enl{\`e}ve l'{\'e}chelle: {\'E}l{\'e}ments
  d'analyse des aspects multiniveaux dans la simulation {\`a} base d'agents.
\newblock {\em Revue d'Intelligence Artificielle}, 24(5):625--648.

\bibitem[Gil-Quijano et~al., 2012]{Gil-Quijano:2012}
Gil-Quijano, J., Louail, T., and Hutzler, G. (2012).
\newblock From biological to urban cells: Lessons from three multilevel
  agent-based models.
\newblock In Desai, N., Liu, A., and Winikoff, M., editors, {\em Principles and
  Practice of Multi-Agent Systems}, volume 7057 of {\em Lecture Notes in
  Computer Science}, pages 620--635. Springer.

\bibitem[Kesting et~al., 2010]{Kesting:2010}
Kesting, A., Treiber, M., and Helbing, D. (2010).
\newblock Enhanced intelligent driver model to access the impact of driving
  strategies on traffic capacity.
\newblock {\em Philosophical Transactions of the Royal Society A: Mathematical,
  Physical and Engineering Sciences}, 368(1928):4585--4605.

\bibitem[Magne et~al., 2000]{Magne:2000}
Magne, L., Rabut, S., and Gabard, J. (2000).
\newblock Towards an hybrid macro-micro traffic flow simulation model.
\newblock {\em Proceedings of the INFORMS Salt Lake City String 2000
  Conference}.

\bibitem[Mammar and Haj-Salem, 2006]{Mammar:2006}
Mammar, S.and~Lebacque, J. and Haj-Salem, H. (2006).
\newblock Hybrid model based on second-order traffic model.
\newblock {\em 85th Annual Meeting of Transportation Research Board},
  1(06-2160).

\bibitem[Michel, 2007a]{Michel:2007}
Michel, F. (2007a).
\newblock The {IRM4S} model: the influence/reaction principle for multiagent
  based simulation.
\newblock In {\em Proc. of 6th Int. Conf. on Autonomous Agents and Multiagent
  Systems (AAMAS 2007)}, pages 1--3.

\bibitem[Michel, 2007b]{Michel:2007a}
Michel, F. (2007b).
\newblock Le mod\`ele {IRM4S}. de l'utilisation des notions d'influence et de
  r\'eaction pour la simulation de syst\`emes multi-agents.
\newblock {\em Revue d'Intelligence Artificielle}, 21:757--779.

\bibitem[Morvan, 2013]{Morvan:2013}
Morvan, G. (2013).
\newblock Multi-level agent-based modeling - a literature survey.
\newblock {\em CoRR}, abs/1205.0561.

\bibitem[{Morvan} and {Jolly}, 2012]{Morvan:2012b}
{Morvan}, G. and {Jolly}, D. (2012).
\newblock {Multi-level agent-based modeling with the Influence Reaction
  principle}.
\newblock {\em CoRR}, abs/1204.0634.

\bibitem[Morvan and Kubera, 2014a]{Morvan:2014a}
Morvan, G. and Kubera, Y. (2014a).
\newblock On time and consistency in multi-agent-based simulations.
\newblock Working paper.

\bibitem[Morvan and Kubera, 2014b]{Morvan:2014}
Morvan, G. and Kubera, Y. (2014b).
\newblock {SIMILAR}: Simulations with multi-level agents and reactions.
\newblock Working paper.

\bibitem[Morvan et~al., 2011]{Morvan:2011}
Morvan, G., Veremme, A., and Dupont, D. (2011).
\newblock {IRM4MLS}: the influence reaction model for multi-level simulation.
\newblock In Bosse, T., Geller, A., and Jonker, C., editors, {\em
  Multi-Agent-Based Simulation XI}, volume 6532 of {\em Lecture Notes in
  Artificial Intelligence}, pages 16--27. Springer.

\bibitem[Picault and Mathieu, 2011]{Picault:2011}
Picault, S. and Mathieu, P. (2011).
\newblock An interaction-oriented model for multi-scale simulation.
\newblock In Walsh, T., editor, {\em Twenty-Second International Joint
  Conference on Artificial Intelligence}, pages 332--337, Barcelona, Spain.
  AAAI Press.

\bibitem[Poschinger et~al., 2002]{Poschinger:2002}
Poschinger, A., Kates, R., and Keller, H. (2002).
\newblock Coupling of concurrent macroscopic and microscopic traffic flow
  models using hybrid stochastic and deterministic disaggregation.
\newblock {\em Transportation and Traffic Theory for the 21st century}.

\bibitem[S{\'e}tra, 2010]{Setra:2010}
S{\'e}tra (2010).
\newblock Identification et localisation sur le reseau routier national.
\newblock Technical report, S{\'e}tra, 110 rue de Paris 77171 Sourdun - France.

\bibitem[Sewall et~al., 2011]{Sewall:2011}
Sewall, J., Wilkie, D., and Lin, M.~C. (2011).
\newblock Interactive hybrid simulation of large-scale traffic.
\newblock {\em ACM Transaction on Graphics (Proceedings of SIGGRAPH Asia)},
  30(6).

\bibitem[Soyez et~al., 2013]{Soyez:2013}
Soyez, J.-B., Morvan, G., Dupont, D., and Merzouki, R. (2013).
\newblock A methodology to engineer and validate dynamic multi-level
  multi-agent based simulations.
\newblock In {\em Multi-Agent-Based Simulation XIII}, volume 7838 of {\em
  Lecture Notes in Artificial Intelligence}, pages 130--142. Springer.

\bibitem[Vo, 2012]{Vo:2012b}
Vo, A. (2012).
\newblock {\em An operational architecture to handle multiple levels of
  representation in agent-based models}.
\newblock PhD thesis, Universit{\'e} Paris VI.

\bibitem[Vo et~al., 2012a]{Vo:2012a}
Vo, D.-A., Drogoul, A., and Zucker, J.-D. (2012a).
\newblock An operational meta-model for handling multiple scales in agent-based
  simulations.
\newblock In {\em International Conference on Computing and Communication
  Technologies, Research, Innovation, and Vision for the Future (RIVF)}, pages
  1--6, Ho Chi Minh City. IEEE.

\bibitem[Vo et~al., 2012b]{Vo:2012}
Vo, D.-A., Drogoul, A., Zucker, J.-D., and Ho, T.-V. (2012b).
\newblock A modelling language to represent and specify emerging structures in
  agent-based model.
\newblock In Desai, N., Liu, A., and Winikoff, M., editors, {\em Principles and
  Practice of Multi-Agent Systems}, volume 7057 of {\em Lecture Notes in
  Computer Science}, pages 212--227. Springer.

\end{thebibliography}
%}

\end{document}